\documentclass[aps,prl,showpacs,superscriptaddress,groupedaddress,longbibliography]{revtex4-1}  
\usepackage{graphicx}  
\usepackage{dcolumn}   
\usepackage{bm}        
\usepackage{amssymb}   

\pdfoutput=1
\usepackage{color}
\usepackage{subfigure,hyperref}
\usepackage{lineno}

\usepackage{latexsym,amsmath,verbatim}

\begin{document}

\title{Creep failure of hierarchical materials}
\author{Mahshid Pournajar$^1$, Paolo Moretti $^1$, Seyyed Ahmad Hosseini$^1$ and Michael Zaiser$^{1}$}
\affiliation{$^1$Dept. of Materials Science, WW8-Materials Simulation, FAU Universit\"at Erlangen-N\"urnberg, Dr.-Mack-Stra{\ss}e 77, 90762 F\"urth, Germany}

\begin{abstract} 
Creep failure of hierarchical materials is investigated by simulation of beam network models. Such models are idealizations of hierarchical fibrous materials where bundles of load-carrying fibers are held together by multi-level (hierarchical) cross-links. Failure of individual beams is assumed to be governed by stress-assisted thermal activation over local barriers, and beam stresses are computed by solving the global balance equations of linear and angular momentum across the network. Disorder is mimicked by a statistical distribution of barrier heights. Both initially intact samples and samples containing side notches of various length are considered. Samples with hierarchical cross-link patterns are simulated alongside reference samples where cross-links are placed randomly without hierarchical organization. The results demonstrate that hierarchical patterning may strongly increase creep strain and creep lifetime while reducing the lifetime variation. This is due to the fact that hierarchical patterning induces a failure mode that differs significantly from the standard scenario of failure by nucleation and growth of a critical crack. Characterization of this failure mode demonstrates good agreement between the present simulations and experimental findings on hierarchically patterned paper sheets. 
\end{abstract}


\maketitle

\section{Introduction}

Hierarchical materials have microstructures whose geometrical features repeat on different length scales in a self-similar fashion. Biological materials provide compelling examples: Materials such as nacre, bivalve shell, bone, and tendon collagen exhibit multi-level hierarchical microstructures which extend from the nanoscale to the macroscale. Such microstructures serve a dual purpose: By efficiently bridging from the nanoscale, on which the smallest hierarchical features are found, to the macroscale, they directly harness nanoscale size effects, which endow the smallest features with increased strength as compared to macroscopic structures consisting of the same material \cite{fratzl2007nature}. At the same time, it has been suggested that hierarchical organization may delay or prevent the spreading of flaws driven by local stress concentrations e.g. at crack tips, which often controls failure of non-hierarchical materials. This may happen because, as shown by \cite{gao2006application}, hierarchical architecture may extend the fracture process zone in such a manner that it ultimately reaches the specimen dimensions and stress/strain concentrations are avoided. As a consequence, hierarchical materials may exhibit remarkable stiffness, good toughness, and high strength, even when they are made of brittle constituents   \citep{Sun2012, Jiao2015, Gao2006, Rho1998, Gautieri2011, Sen2011}. 

Numerical modeling has been envisaged as a tool for designing hierarchical structures and predicting their properties. Simple modelling approaches such as fiber bundle models have been extended to hierarchical materials \cite{Biswas2019} but cannot adequately represent crack-tip stress concentrations which are an essential aspect in the failure of materials by crack propagation. Spatial stress concentrations and crack propagation can be described by fuse and beam network models. As an example, \citet{Moretti2018} used a random fuse model (RFM) to comparatively study failure behaviour of materials with and without hierarchical microstructure. The failure behaviour of hierarchical materials, as well as the resulting super-rough crack morphology, was found to differ significantly from the behavior of non-hierarchical ones. \citet{Hosseini2021} introduced a hierarchical version of a beam network model (BNM) to simulate hierarchically patterned materials. The findings of these authors show that the failure of hierarchically structured materials is caused by local damage nucleation followed by damage percolation rather than critical crack propagation, a finding which holds for both two-dimensional \citep{Hosseini2021} and three-dimensional \citep{hosseini2023enhanced} structures. In both cases, crack profiles show large deflections as failure arises from the coalescence of widely separated micro-cracks. \citet{Zaiser2022} conducted both simulations and experiments on two-dimensional quasi-brittle sheets loaded along a single axis. It was demonstrated that hierarchical patterning considerably increased crack propagation resistance both in terms of the peak stress and the work of failure of pre-cracked samples. The differences in failure mode between hierarchical and non-hierarchically patterned 2D material were further investigated by \citet{pournajar2023failure}, who studied failure precursors in patterned paper sheets under both displacement-controlled and creep loading. Failure precursors were characterized in terms of acoustic emission and digital image correlation to map the evolution of strain patterns in the run-up to failure. The results confirm the idea that hierarchical patterning efficiently mitigates against stress/strain concentrations and promotes a diffuse mode of failure that is insensitive to the existence of even large flaws. 

The hierarchical structures considered in the above mentioned works can be alternatively envisaged as assemblies of laterally connected load carrying fibers which are bundled into hierarchically nested ``modules'', or as solid sheets (in 3D: blocks) of a matrix material that are divided into modules by load-parallel gaps and that exhibit a power-law distribution of gap lengths (in 3D: gap areas). From the perspective of composite mechanics, such structures can be considered as limit cases of ``calcite-like'' hierarchical composites as studied by Buehler and co-workers \cite{Sen2011,mirzaeifar2015defect} where weak and compliant, lamellar inclusions are embedded into a hard and brittle matrix in a hierarchical manner: In our work the inclusions have zero elastic modulus and near-zero volume fraction as they correspond to the gaps, whereas the matrix corresponds to the matrix material. 

Most simulation studies of failure of hierarchical materials have focused on load-driven failure, notably under uni-axial loading conditions, where a load or displacement is imposed on the system and increased until the system loses connectivity. Typical of such scenarios, the process of damage accumulation and ultimate failure does not depend on loading rate, i.e. the material behavior is time independent. In real materials, however, an important failure pathway is that of creep, or sub-critical failure, where the system accumulates damage and approaches failure over an extended period of time. Here we thus investigate creep failure of hierarchical materials, i.e. the time evolution of damage and fracture under conditions of constant sub-critical load.

\section{Methods}

\subsection{Construction and geometrical properties of the studied beam networks}

The construction method which we use has been described in detail elsewhere \cite{Moretti2018,Hosseini2021}. Here, we only briefly summarize the main features. In all cases, we start out from a simple square lattice of beams where the loading axis is aligned with one of the cubic lattice directions. The size of the lattice is $L$ which we
take to be a power of 2, $L=2^n$, where $n$ will define the number of hierarchical levels. There are $L^2$ load-carrying beams in loading (henceforth 'vertical') direction forming $L$ load-carrying columns, and $L^2$ beams that are initially unloaded ('cross links') in lateral ('horizontal') direction forming $L$ rows. The points at which the beams are connected are denoted as nodes; the endpoints of the load carrying columns are denoted as upper and lower boundary nodes. In all cases, loading is carried out by keeping the bottom boundary nodes fixed and displacing the top boundary nodes rigidly upwards such as to create a prescribed total load. Periodic boundary conditions are applied in lateral direction. From this basic set-up, different variants are constructed by removal of beams while the node locations are always kept the same.

\begin{figure}[tb]
\includegraphics[width=0.8\textwidth]{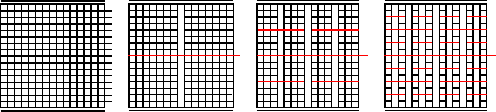}
\includegraphics[width=0.7\textwidth]{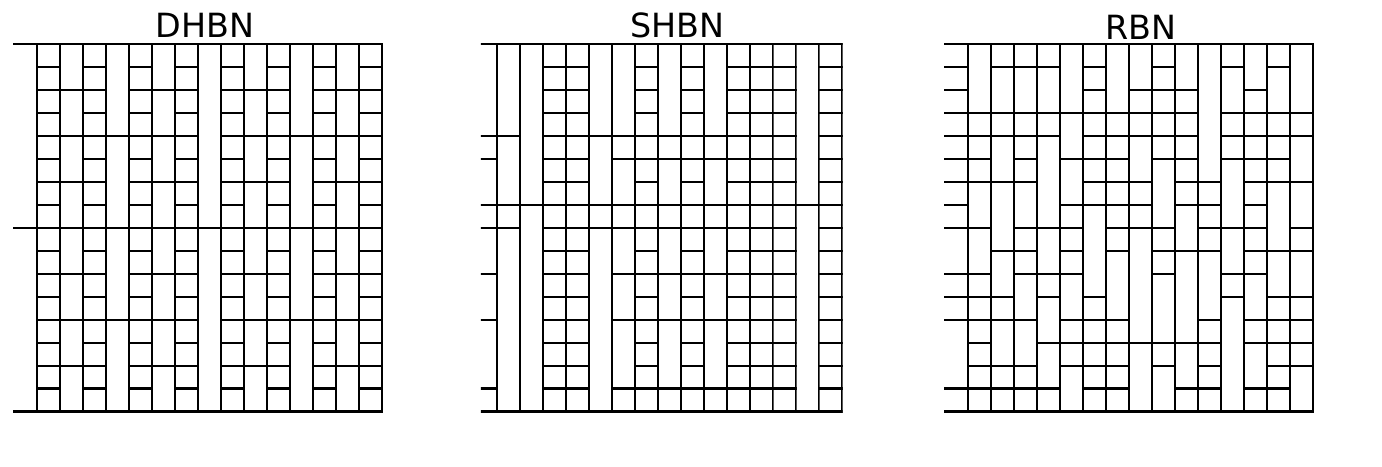}
\caption{Top: iterative 'top-down' construction of a deterministic hierarchical beam network: a beam network is divided by load-parallel cuts (removal of vertically adjacent cross links) into four highest-level modules. These form two groups of two modules loaded in series, connected by a system spanning connector (the row of cross links shown in red); next, each of the four modules is again divided into four lower-level modules plus a module-spanning lateral connector, etc.; Bottom: Network variants as discussed in the text.}
\label{fig:networkvariants}
\end{figure}

To construct deterministic hierarchical beam networks (DHBN), a hierarchical structure is obtained by removing cross links such as to create gaps which recursively sub-divide the structure into load-carrying modules of decreasing order as illustrated in Figure \ref{fig:networkvariants}, top.
This construction leads to a hierarchical pattern of modules separated by gaps, and of connecting cross links. In addition to DHBN, we consider a randomized variant, so called stochastic hierarchical beam networks, SHBN, constructed by first randomly reshuffling the columns and then the resulting rows. Furthermore, as non hierarchical reference structures, random beam networks (RBN) are constructed by removing randomly the same number of cross links as in a DHBN or SHBN structure. The different network variants are illustrated in Figure \ref{fig:networkvariants}, bottom.

In all cases, we may introduce cracks of integer length $a$ into the structure by removing, at a randomly chosen location, a row of $a$ adjacent load carrying beams. 

\subsection{Stress re-distribution}
\label{sec:equations}
Displacement of the boundary nodes induces forces and moments in the beam network. To evaluate these forces we envisage straight, identical beams of unit modulus of elasticity, unit length and square unit cross section which are capable of transmitting axial and shear forces as well as bending moments. The beams are rigidly clamped to the nodes, and each node has two translational degrees of freedom (node displacements $u$ and $v$ along the global $x$ and $y$ axes) and one rotational degree of freedom (rotation angle $\theta$ about the global $z$ axis). The beams are assumed to deform in a linearly elastic manner, their deformation is described using Timoshenko beam theory. Evaluating the resulting balance equations for linear and angular momentum at all nodes in the quasi-static limit and using a small-strain approximation results in a linear system of equations for the nodal degrees of freedom, whose solution gives us the local forces $F_i$ and bending moments $M_i$ acting on all beams (for details see citet{Hosseini2021}).

\subsection{Failure criterion}
\label{sec:failure}
Beams deform in a linearly elastic manner and undergo time-dependent failure where the failure time is controlled by the maximum deviatoric stress magnitude acting on the beam. We define the maximum deviatioric stress acting on the beam connecting nodes $i$ and $j$ as
\begin{equation}\label{eq:failurestress}
	\sigma_{ij} = \sqrt{ \Big( \dfrac{F_{i}n_i}{A}+{\rm max}(|M_{i}|,|M_{j}|)\dfrac{y_{\rm max}}{I} \Big) ^ {2}+3 \Big( \dfrac{Q_{i}}{A} \Big) ^ {2} } 
\end{equation} 
where $n_i$ indicates the outward normal direction of the beam end surface connecting to node $i$, $F_i$ is the normal force which can be tensile ($F_i n_i > 0$) or compressive ($F_i n_i < 0$), $Q_i$ is the shear force, and $M_i$ is the moment acting on this surface. A derivation of Eq. (\ref{eq:failurestress}) based on the Maximum Distortion Energy of Failure (von Mises) criterion was given by \citet{Hosseini2021}. From the deviatoric stress, the beam failure rate is calculated under the assumption that it is controlled by a thermally activated process: 
\begin{equation}\label{eq:failurerate}
	\nu_{ij} = \nu_0 \exp\left[-\frac{H_{ij} - \sigma_{ij}V_{\rm a}}{k_{\rm B} T}\right] =  \nu_0 \exp\left[-\frac{t_{ij} - \sigma_{ij}}{\Theta}\right] 
\end{equation} 
where $\nu_0$ is a characteristic attempt frequency, $H_{ij}$ is the activation barrier for beam failure, $t_{\rm ij}=H_{ij}/V_{\rm a}$ is the corresponding beam failure stress, $V_{\rm a}$ is an activation volume, and $\Theta = k_{\rm B} T/(V_{\rm a}$ is the characteristic thermal energy therein. Eq. (\ref{eq:failurerate}) holds as long as the beam stress is below the beam failure stress, $\sigma_{ij} < t_{ij}$ (subcritical failure), if on the other hand $\sigma_{ij} \ge t_{ij}$, then failure occurs instantaneously. 

Mimicking material heterogeneity, beam failure stresses $t_{ij}$ are randomly assigned based on a Weibull probability distribution function with probability density

\begin{equation}\label{eq:weibull}
	p(t_{ij}) = \dfrac{\beta}{\eta} \Big( \dfrac{t_{ij}}{\eta} \Big) ^ {\beta - 1} \exp\left(-\Big( \dfrac{t_{ij}}{\eta} \Big)^\beta \right)
\end{equation}
where $\beta  > 0$ and $\eta  > 0$ are the shape and scale parameters of the distribution, respectively. In this paper we normalize all stresses by the mean failure threshold stress and adjust the shape parameter $\eta$ such as to have a fixed mean value $\langle t_{ij} \rangle = 1.0$. The shape parameter $\beta$ may be varied to implement different degrees of disorder.

\subsection{Simulation protocol}
Simulations are performed to mimic creep conditions, i.e., deformation under a stationary temporally constant global load $\sigma_{\rm c}$. However, as in experiment, imposing this load requires an initial, non stationary loading stage.
\subsubsection{Initial loading}
The simulation starts with an initial loading stage where the external axial displacement is increased until the global stress level $\sigma$ (evaluated as the sum of axial beam forces on the top nodes, divided by $L$) reaches the prescribed level $\sigma = \sigma_{\rm c}$ {\em or} a beam breaks because $t_{ij} \le \sigma_{ij}$. If beam failure occurs during initial loading, the displacement is then kept fixed while load is re-distributed which may trigger further instantaneous failures. Load re-calculation is repeated iteratively in such a way that one beam is removed in every iteration, until all beams are in a subcritical state ($\sigma_{ij} > t_{ij})$; this sequence of failures in general leads to a stress relaxation. The axial displacement is then again increased until either the prescribed creep stress is reached or beams fail, and this is repeated until the system reaches a sub-critical state at $\sigma = \sigma_{\rm c}$ or the system becomes disconnected; in the latter case, we terminate the simulation and note instantaneous system failure during initial loading. 

\subsubsection{Creep behavior}

Once the system has settled into a sub-critical state, we use a Kinetic Monte Carlo Algorithm to determine the location of the next beam failure. We count the number of thermally activated failures by the discrete index $n_T$. To identify the failing beam, we evaluate the failure rates $\nu_{ij}$ for all remaining intact beams and choose one of them with probability $p_{ij} = \nu_{ij}/\nu_{\rm tot}$ where $\nu_{\rm tot} = \sum_{ij} \nu_{ij}$. This beam is removed from the system, $n_T$ is increased by on, and time is increased by $\Delta t(n_T)$ which we take to be an exponentially distributed random variable of average $\nu_{\rm tot}^{-1}$. 

Subsequent to a thermally activated beam failure, the stresses on all beams are re-computed, and beams where $\sigma_{ij}> t_{ij}$ are removed instantaneously. The external displacement is then adjusted to maintain the global stress at $\sigma = \sigma_{\rm c}$, and the process is repeated until either all remaining beams are in a sub-critical state or the system disconnects. The number of beams that fail instantaneously after the $n_T$th  thermally activated failure is denoted the avalanche size $s(n_T)$, the displacement after the avalanche is $d(n_T) = L \epsilon(N_T)$ where  $\epsilon(N_T)$ is the total creep strain, and the avalanche time is $\sum_{n_T}\Delta t(n_T)$. 

After an avalanche has terminated, we again use a Kinetic Monte Carlo step to determine the next thermally activated beam failure, and this sequence is repeated until the system globally disconnects. The corresponding time is called the creep lifetime $t_{\rm c}$. 

\section{Results}

\subsection{Creep curves and time to failure}

\begin{figure}[p]
\includegraphics[width=\textwidth]{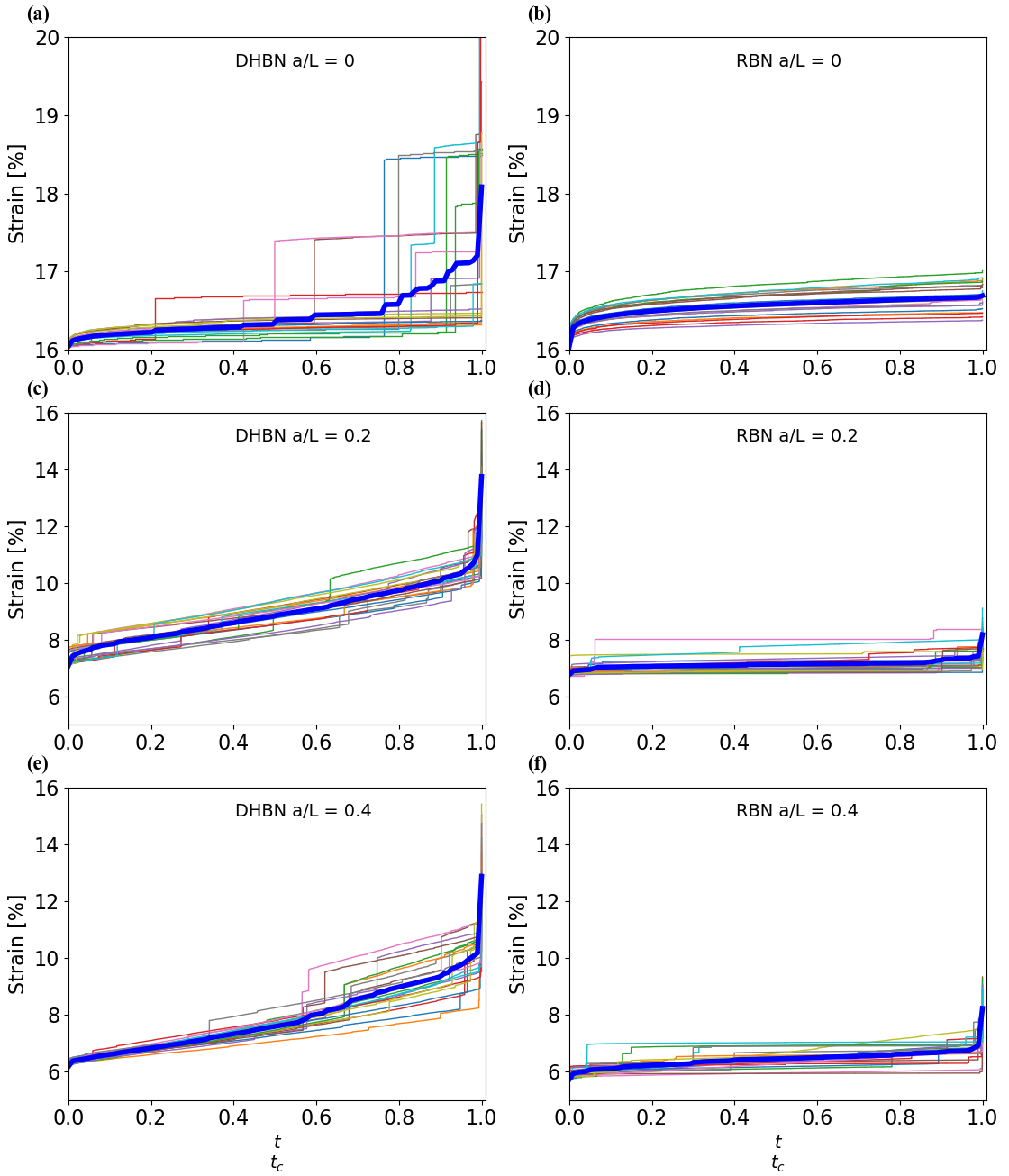}
\caption{Creep strain vs. normalized time curves for structures of size $L=512$; left: deterministic hierarchical beam network (DHBN)- a)un-notched, c)notch length = 0.2L, e) notch length = 0.4L; right: random beam network (RBN)- b)un-notched, d)notch length = 0.2L, f) notch length = 0.4L; each plot shows results of 20 simulated samples, the blue lines represent the average creep strain curves; for each crack length, the applied stress amounts to 80\% of the mean RBN failure stress at the respective crack length: $(\sigma/\langle \sigma_{\rm f}\rangle)$ = 0.04, 0.06, 0.16 for $a = 0.4L, 0.2L, 0$ respectively; for other parameters, see text.}
\label{fig:creep-strain-time}
\end{figure}

Figure \ref{fig:creep-strain-time} shows typical creep curves (creep strain vs time) obtained for DHBN and RBN samples with different initial crack lengths. In all simulations, temperature was set to $\Theta = 0.01$ and the Weibull parameter of the barrier height distribution was $\beta = 4$. In the plots, the time axis is normalized by the failure time, which allows to average curves pertaining to different samples to obtain average creep curves (blue lines in Figure \ref{fig:creep-strain-time}). 

With exception of the un-cracked RBN, all samples show typial three-stage creep curves. During initial loading, an instantaneous strain builds up which is almost identical for hierarchical DHBN and non-hierarchical RBN samples. This is followed by a regime of decreasing creep rate (creep stage I), then an extended regime of almost constant slope (creep stage II) and an accelerating regime in the run-up to failure (creep stage III). 

Stage II shows clear differences between DHBN and RBN samples. First of all, the overall strain that accumulates in Stage II is significantly higher in the hierarchical samples. Second, the deformation process of hierarchical samples is puctuated by large avalanches, where the rate of deformation suddenly jumps upwards and then relaxes back to a low level. In single samples, one might speak of multi-stage creep, however, the random timing of the strain bursts leads on average to a smooth average response (blue lines in Figure \ref{fig:creep-strain-time}, left) which is still well described as Stage-II creep with a well defined average creep rate. The observation of large strain bursts in the stage-II creep curves of hierarchical samples in our simulations matches experimental observations of \citet{pournajar2023failure}, who reported such bursts in hierarchically patterned paper samples but not in their non-hierarchical counterparts. 

Stage III is less pronounced in RBN samples where the onset of failure is more sudden than in DHBN samples. In uncracked RBN samples this stage is almost absent. These observations match the findings for the same sample types in displacement-controlled uni-axial loading: In RBN samples, the pattern of failure is essentially brittle -- loading is almost elastic up to the critical failure stress, and after the peak load one observes an almost vertical load drop with little or no post-peak activity. These findings match the observation in the creep simulations of a small stage-II creep strain before the onset of softening and almost no strain accumulation after the onset of softening. In DHBN samples, on the other hand, displacement controlled tests are characterized by significantly higher strains-to-failure which accumulate both during loading, where damage accumulates in a series of avalanche-like bursts, and in the form of significant post-peak activity, leading to a much more gradual mode of failure as compared to the RBN samples.

\begin{figure}[thb]
\centering
\includegraphics[width=0.32\textwidth]{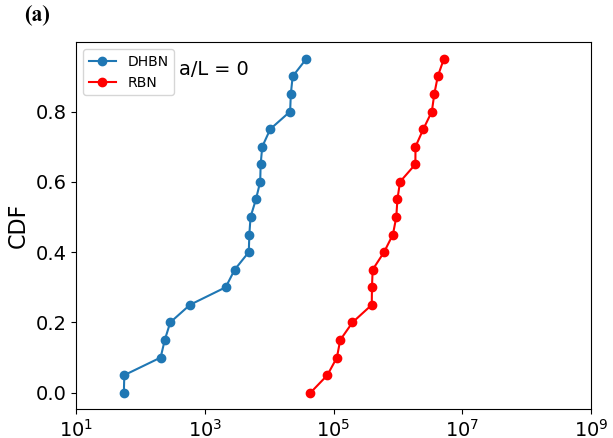}\hfill
\includegraphics[width=0.32\textwidth]{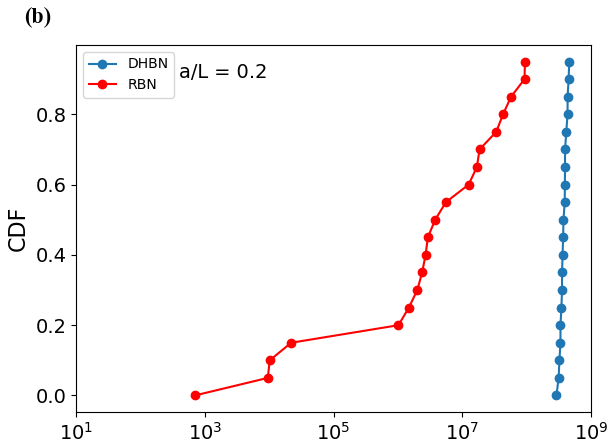}\hfill
\includegraphics[width=0.32\textwidth]{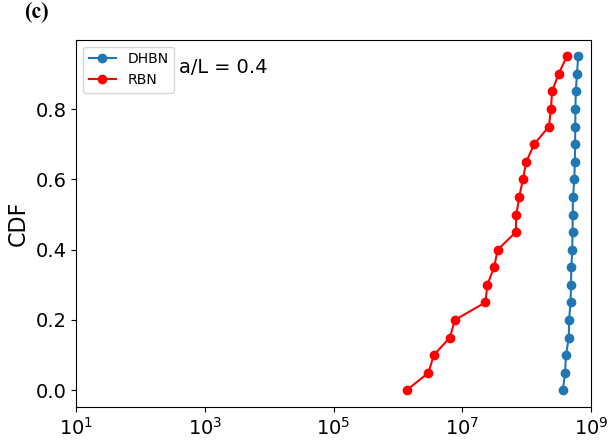}\hfill
\\
\includegraphics[width=0.32\textwidth]{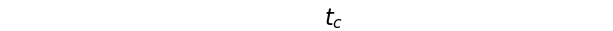}\hfill
\includegraphics[width=0.32\textwidth]{CDFcreep4.png}\hfill
\includegraphics[width=0.32\textwidth]{CDFcreep4.png}\hfill
\caption{Cumulative distribution of creep failure time for different structures of size $L=512$, deterministic hierarchical beam network (DHBN) and random beam network (RBN)- a)un-notched, b)notch length = 0.2L, c) notch length = 0.4L; The constant applied stress on both DHBN and RBN is $80\%$ of the peak stresses of RBN for each crack length, which corresponds to creep stresses in units of the mean beam failure stress $\langle\sigma_{B}\rangle$ of 0.04, 0.06 and 0.16 for systems with crack lengths $a= 0.4L, a= 0.2L$, and $a=0$, respectively. Each plot represents the results of 20 simulations. For other parameters, see text.}
\label{fig:creep-CDF}
\end{figure}

Regarding time to failure, a complex picture emerges. For notched samples, the lifetime of DHBN samples significantly exceeds that of notched RBN samples subjected to the same creep load (Figures \ref{fig:creep-CDF}, (b) and (c)). At the same time, the lifetime scatter is very significantly reduced. Thus, in samples containing macroscopic flaws, hierarchical patterning not only provides enanced flaw tolerance but also ensures a more predictable material response. The picture is different in case of samples without initial macroscopic notch (Figure \ref{fig:creep-CDF}, (a)). In this case, the logarithmic scatter of the lifetimes is comparable for DHBN and RBN samples, while the absolute creep lifetimes of the RBN samples are higher than those of the DHBN samples.

\subsection{Spatial and temporal patterns of damage accumulation}

\begin{figure}[b]
\centering
\includegraphics[width=.8\textwidth]{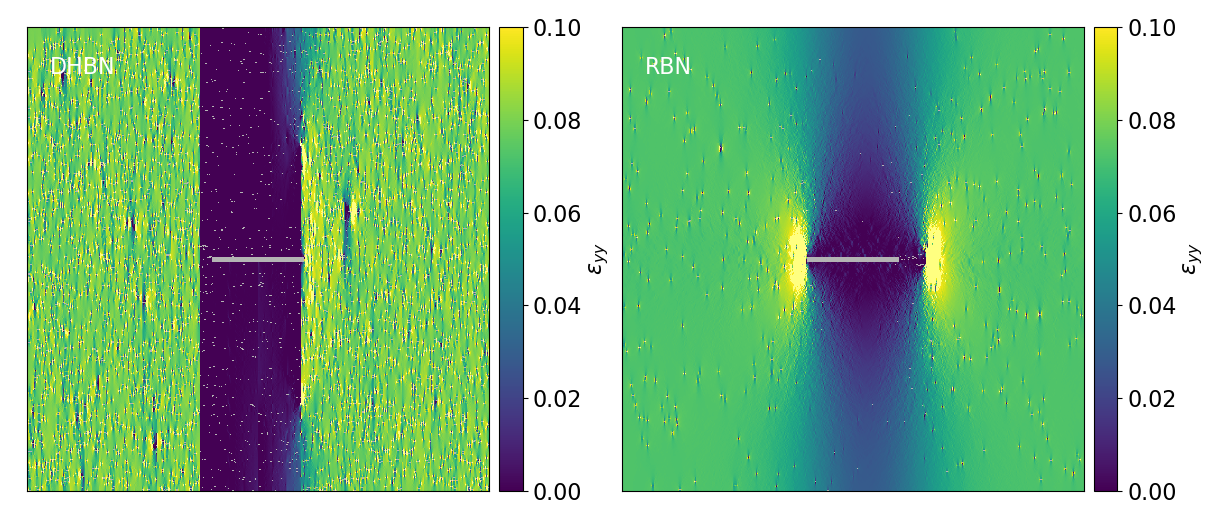}
\caption{Axial strain $\epsilon_{yy}$ map, left: deterministic hierarchical beam network (DHBN), right: random beam network (RBN) structures of size $L=512$- during creep steady state step at 0.5 failure time $t_{c}$ in a creep simulation;  The constant applied stress on both DHBN and RBN is 80\% of the peak stress of RBN which is 0.06 in unit of mean beam failure stress $\langle\sigma_{B}\rangle$ for systems having 0.2L precrack length. The precrack is marked in gray, for other parameters, see text.}
\label{fig:strain-pattern}
\end{figure}

In order to analyze the reasons for the characteristic differences between the behavior of hierarchical and non-hierarchical samples, we look at the spatial and temporal patterns of damage accumulation that emerge in the run-up to creep failure. Figure \ref{fig:strain-pattern} shows typical strain patterns in creep stage II in pre-cracked samples (crack length $a=0.2 L$), at about half the sample lifetime. For interpretation of the patterns it is important to note that, in a homogeneous sample and in absence of creep damage, the imposed creep stress of $\sigma_{\rm c} = 0.06$ would induce a corresponding homogeneous elastic strain of $\epsilon_{yy} = 0.06$. Because of the crack in the sample center, the stress and strain above and below the crack are significantly reduced (shielding) while the strain ahead of the crack tips is enhanced. At the same time, we observe very significant differences between DHBN and RBN samples. In the RBN samples, damage accumulation is confined to two process zones right ahead of the crack tips, whereas in the rest of the sample, deformation remains almost elastic. Damage accumulation at the crack tip leads to a gradual crack advance, and the transition to failure is related to the pre-existing crack becoming supercritical. In DHBN samples, on the other hand, we observe a much higher degree of global damage accumulation. This damage is not localized at the tips of the pre-existing crack but distributed over the entire sample with exception of the shielded regions above and below the crack. The diffuse nature of damage accumulation without supercritical crack propagation allows the hierarchical material to sustain a much higher creep strain. Again these findings are in excellent agreement with the experimental observations reported by \citet{pournajar2023failure} who used digital image correlation to map the strain distribution in paper samples with both hierarchical and non hierarchical cut patterns. 

We next look at the statistics of damage bursts or avalanches occurring during creep deformation. Failure of one beam leads to load being re-distributed onto other beams, and this can give rise to secondary failures, and ultimately to a cascade of beam failures (``avalanches''). While most thermally activated beam failures trigger only very few sequels, the largest avalanches may encompass hundreds of beam failures and are directly visible in form of sudden large strain increments on the creep curves (see Figure \ref{fig:creep-strain-time}). In the following, we define the avalanche size $S$ as the number of instantaneous, ``athermal'' beam failures after any given thermally activated beam failure. 

Avalanches emerging in failure of disordered materials often exhibit power-law statistics (for an overview, see \cite{alava2006statistical}). This is also true for hierarchical materials subject to load-controlled deformation, as found in simulation studies \cite{Moretti2018,Hosseini2021} as well as in experiments \cite{pournajar2023failure}. 

\begin{figure}[t]
\includegraphics[width=0.49\textwidth]{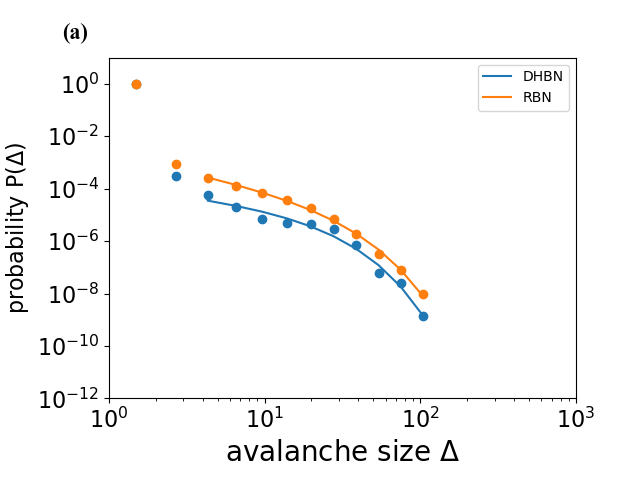}\hfill
\includegraphics[width=0.49\textwidth]{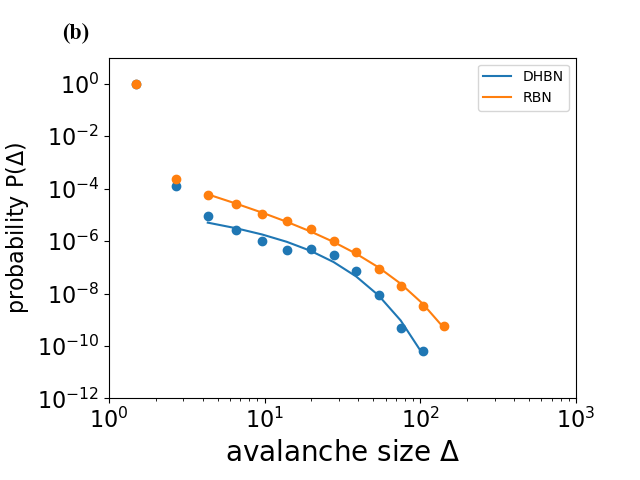}\hfill\\
\includegraphics[width=0.49\textwidth]{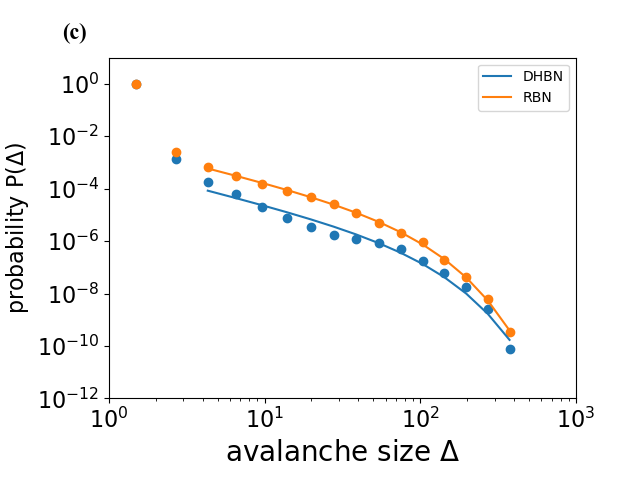}\hfill
\includegraphics[width=0.49\textwidth]{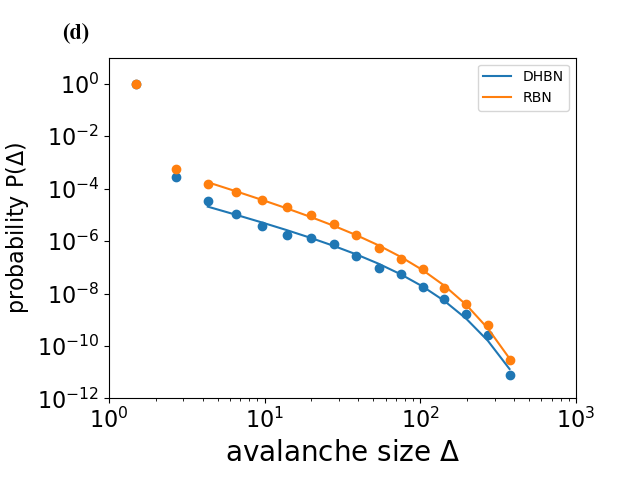}\hfill
\caption{Avalanche size distributions (probability vs number of broken beams) during the different stages of the creep curve.}
\label{fig:avalstat}
\end{figure}

We determine probability distributions $P(\Delta)$ by logarithmic binning, approximating 
\begin{equation}
    P(\Delta) \approx \frac{N(S,\Delta_S)}{N_{\infty} \Delta_S} \quad,\quad N(S,\Delta_S) = N(S_i|(S-\Delta_S/2) < S_i < (S+  \Delta_S/2))
\end{equation}
where the bin sizes $\Delta_S$ increase exponentially with $S$ to ensure adequate statistics in the large-$S$ tails of the distributions. $N_{\infty}$ is the total number of avalanches in the sample. Avalanche statistics was determined from 1000 creep simulations conducted for the same parameters as used in Figure \ref{fig:strain-pattern}. 
Distributions were determined separately for the first $10\%$ of the failure time (roughly corresponding to creep stage I, Figure \ref{fig:avalstat}(a)), the second 80\% (creep stage II, Figure \ref{fig:avalstat}(b) ), and the last 10\% (creep stage III,  Figure \ref{fig:avalstat}(c)), as well as for the total of all avalanches (Figure \ref{fig:avalstat}(d)).

A first conspicuous observation is that the avalanche statistics shows, within the margins of statistical error, no differences between hierarchical and non-hierarchical samples. This is also observed in experiment \cite{pournajar2023failure}, but apparently contradicts the direct observations showing large jumps on the DHBN but not the RBN creep strain vs time curves, which moreover match the observations on experimental creep samples. The reason is simply that the largest avalanches, while conspicuous on the creep curves where they account for a large fraction of the creep strain, carry insufficient statistical weight to shape the statistical distributions. 

Contrary to the findings in load-controlled simulations \cite{Moretti2018,Hosseini2021,hosseini2023enhanced}, the distributions are not well described as truncated power laws. Rather, they are composed of a rapidly decay at small avalanche sizes below $S=5$, an a much slower decaying, long tail in the form of a stretched exponential which extends, in creep stages I and II, to maximum avalanche sizes of about $S=100$. In creep stage III, the maximum avalanche sizes are larger and range up to $S\approx 600$. The total distribution is practically identical to the distribution of stage-III avalanches (Figure \ref{fig:avalstat}(d)), for the simple reason that most avalanches occur in creep stage III.

\section{Discussion and Conclusions}

Our simulation study shows characteristic differences in the creep behavior of hierarchical beam network samples as compared to non-hierarchical reference samples. These differences corroborate, for the case of creep deformation, the idea that hierarchical patterning is a powerful means to enhance the flaw tolerance of  materials, by enhancing the creep lifetime in presence of flaws while reducing the lifetime scatter. This beneficial effect arises from the strong strain de-localization induced by the hierarchical patterning, which reduces crack-tip stress concentrations and reduces statistical scatter by distributing damage accumulation evenly across the sample, thus mitigating against the effect of strength fluctuations near the crack tip which otherwise have a strong impact on sample lifetime. These findings are in good agreement with the experimental data reported by Pournajar et. al. \cite{pournajar2023failure}. At the same time, it must be noted that hierarchical patterning is not always beneficial; in samples without initial notches the creep lifetime is reduced, rather than enhanced, in hierarchical as compared to non hierarchical reference samples. 

In line with experimental studies, our simulations indicate that the statistics of avalanche precursors to creep failure is not strongly influenced by hierarchical patterning. In the context of experimental data reported by \citet{pournajar2023failure}, this could be understood by noting that the avalanche size distributions took the shape of truncated power laws $p(\Delta) \propto \Delta^{-\delta}$ with an avalanche exponent close to $\delta = 1.5$, typical of mean-field behavior. However, the present simulation results are inconsistent with mean-field behavior and further work is needed to understand the nature and statistics of avalanches in the present model.

%

\section{Acknowledgements}
The authors acknowledge financial support from the H2020-MSCA-RISE-2016 program, Grant No. 734485 “Fracture Across Scales and Materials, Processes and Disciplines (FRAMED). M.P. and M.Z. also acknowledge funding by DFG under grant No. Za 171-9/3. 

\section{Author contributions}
M.P. performed the numerical simulations with support if P.M., using program code provided by S.A.H. 
M.P. analyzed the simulation data. 
M.Z. drafted the manuscript. All authors edited and approved the manuscript final version. 

\section{Additional information}
The authors declare no competing interests.

\end{document}